\documentclass[fleqn,usenatbib]{mnras}
\usepackage{newtxtext,newtxmath}
\usepackage[T1]{fontenc}
\usepackage{ae,aecompl}
\usepackage{lastpage}
\usepackage[normalem]{ulem}
\usepackage{graphicx}	
\usepackage{amsmath}	
\usepackage{array, booktabs, makecell} 
\usepackage[normalem]{ulem}
\usepackage{comment}

\usepackage{verbatim} 

\newcommand{\MAXI}{{MAXI\, J1820+070\,}\,}
\newcommand{\MAX}{{J1820\,}\,}

\title[Optical/X-ray QPOs in MAXI J1820+070]{Synchronous X-ray/Optical QPOs from the Black Hole LMXB MAXI J1820+070}

\author[J K Thomas et al.]{Jessymol K. Thomas,$^{1}$\thanks{E-mail: jessy@saao.ac.za}
David A. H. Buckley$^{1,2,3}$,
Philip A. Charles$^{4}$, John A. Paice$^{4,5}$,
\newauthor Stephen B. Potter$^{1,6}$,  James F. Steiner$^{7}$, Jean-Pierre Lasota$^{8,9}$, P. Gandhi$^{4}$, Lian Tao$^{10}$, 
\newauthor Xiang Ma$^{10}$, Yi-Jung Yang $^{10}$,
 Youli Tuo$^{10}$, Shuang-Nan Zhang$^{10}$ \\ \\
$^{1}$ South African Astronomical Observatory, Observatory Road, Observatory, 7925, Cape Town, South Africa\\
$^{2}$Department of Astronomy, University of Cape Town, Private Bag X3, Rondebosch 7701, South Africa\\
$^{3}$Department of Physics, University of the Free State, PO Box 339, Bloemfontein 9300, South Africa\\
$^{4}$Department of Physics $\&$ Astronomy, University of Southampton, Southampton SO17 1BJ, UK\\
$^{5}$Inter-University Centre for Astronomy and Astrophysics, Pune, Maharashtra 411007, India\\
$^{6}$Department of Physics, University of Johannesburg, PO Box 524, Auckland Park 2006, South Africa\\ 
$^{7}$Center for Astrophysics | Harvard \& Smithsonian, 60 Garden St. Cambridge, MA 02138, USA\\
$^{8}$ Institut d’Astrophysique de Paris, CNRS et Sorbonne Université, UMR 7095, 98bis Bd Arago, 75014 Paris, France\\
$^{9}$ Nicolaus Copernicus Astronomical Center, Polish Academy of Sciences, Bartycka 18, 00-716 Warsaw, Poland\\ 
$^{10}$Key Laboratory of Particle Astrophysics, Institute of High Energy Physics, Chinese Academy of Sciences,19B Yuquan Road, Beijing 100049, People's Republic of China \\
\\ \\ 
}
 
\date{Accepted XXX. Received YYY; in original form ZZZ}

\pubyear{2020}

\begin{document}
\label{firstpage}
\pagerange{\pageref{firstpage}--\pageref{lastpage}}
\maketitle


\begin{abstract}
We present high-speed optical photometry from SAAO and SALT on the black hole LMXB \MAXI (ASSASN-18ey), some of it simultaneous with NICER, {\it{Swift}} and \textit{Insight}-HXMT X-ray coverage. We detect optical Quasi-Periodic Oscillations (QPOs) that move to higher frequencies as the outburst progresses, tracking both the frequency and evolution of similar X-ray QPOs previously reported. Correlated X-ray/optical data reveal a complex pattern of lags, including an anti-correlation and a sub-second lag that evolve over the first few weeks of outburst. They also show correlated components separated by a lag equal to the QPO period roughly centered on zero lag, implying that the inter-band variability is strongly and consistently affected by these QPOs at a constant phase lag of roughly $\pm\pi$. The synchronisation of X-ray and optical QPOs indicates that they {\it must} be produced in regions physically very close to each other; we thus propose that they can be explained by a precessing jet model, based on analogies with V404 Cyg and MAXI J1348-630.
\end{abstract}



\section{Introduction}\label{sec:1}

A prominent feature of Black Hole X-ray Binaries (BHXBs) in outburst is the presence of quasi-periodic oscillations (QPOs), short time-scale modulations exhibited in the X-ray light-curves. Though the QPOs in BHXB transients have been studied extensively, their origin, evolution and physical nature remain controversial. 

Both X-ray and optical QPOs have been seen in the bright BHXB MAXI J1820+070 (ASASSN-18ey) (hereafter J1820), discovered in March 2018 in X-rays by \citet{ATel11399} and then as an optical transient by \citet{Denisenko18}. X-ray QPOs from \MAX have been found by \textit{INTEGRAL}, who detected a low frequency, $0.04\pm 0.01$ Hz QPO with the JEM-X monitor in late March 2018 \citep{Atel11488}, and AstroSat confirmed an X-ray QPO of 0.0477 Hz \citep{Astrosat}. Also, a QPO at $\sim$0.06 Hz was reported in April from {\textit{Swift}} X-ray power spectra \citep{ATel11510}. 

The X-ray QPO properties of \MAX from NICER observations have been presented in detail by \cite{Stiele2020}, showing how their frequencies evolve smoothly from $\sim$ 0.03 Hz on day 10 to $\sim$10 Hz at the time of the hard to soft state transition. For convenience we adopt the same time reference as used by \cite{Stiele2020} (i.e. day 0 is 2018 Mar 11 0:00 UT = MJD 58188), just before the first triggering of X-rays from J1820 \citep{ATel11399}. \textit{Insight}-HXMT has detected 1--200\,keV X-ray OPQs with frequency gradually evolving from 0.02\,Hz to 0.65\,Hz in the hard and intermediate states \citep{Ma2021}. X-ray QPOs from NuSTAR observations were also reported \citep{Buisson_NuSTARJ1820_1_2019}. Furthermore, a QPO of 0.0495 Hz was reported from optical photometry with the Lijiang 2.4m telescope of Yunnan observatories by \cite{ATel11510} on March 30, 2018. The INAF-Astronomical Observatory of Padova also detected a 0.128\,Hz optical QPO with IFI+Iqueye and Aqueye on April 18 and 19, and two more QPO-like features (0.268 Hz and 0.151 Hz) were detected on June 9--10, using the same instruments \citep{Zampieri2019}. 

The number of QPOs detected and their characteristic frequency range depends on the system properties. QPO frequencies $<$1Hz have been seen in Cyg X-1 \citep{Pottschmidt2003} and low frequency QPO-like features in the range $\sim$0.03Hz – 0.05 Hz from hard state observations of XTE J1752-223 \citep{MUNOZ2010}. Similar QPOs were also detected from GX\,339-4 in 2004 \citep{Motta2011}.

We find low frequency optical QPOs in our J1820 hard state data, which we discuss in Section \ref{sec:opticalQPOs} and show in Section \ref{sec:opticalXrayQPOs} that they closely track the previously reported X-ray QPOs. We used  SALT and the SAAO 1\,m to obtain high-speed photometry at various times in the outburst of J1820. These data were cross-correlated (section \ref{sec:CCF}) with  simultaneous X-ray observations from NICER, {\it{Swift}} and \textit{Insight}-HXMT. Our discussion and conclusions are presented in Section 4. For further background information on J1820 together with our long-term variability study see \citet{Thomas21} (hereafter Paper I).

\section{Observations}\label{sec:2}
The optical observations of J1820 used here were carried out with the Southern African Large Telescope (SALT) and the South African Astronomical Observatory (SAAO) 1\,m telescope. There are extensive X-ray observations of J1820 throughout the entire outburst with NICER, {\it{Swift}} and \textit{Insight}-HXMT, some of which have already appeared in the literature, and which we use here to compare with our optical results.

\subsection{SAAO/SHOC}\label{sec:2.1} 

The high speed photometry of J1820 was performed with the SAAO 1\,m for 20 nights, from 25 March to 29 September 2018. The Sutherland High Speed Optical Camera (SHOC, \citealt{2013SHOC}), using a frame-transfer EM-CCD (Andor iXon888), mounted on the 1\,m, was used with a clear filter and exposure times of 200\,ms. The light curves derived from the images were produced from differential aperture photometry using the SHOC data reduction pipeline\footnote{\url{https://shoc.saao.ac.za/Pipeline/SHOCpipeline.pdf}}. We used GSC 00444-02282 from SIMBAD as a comparison star. 

\subsection{SALT- RSS/SALTICAM}\label{sec:2.2}

Photometry of J1820 was also performed with SALT during 5 days in June and July 2018. Observations on 13 June and 7 July used SALT's imaging spectrograph (the Robert Stobie Spectrograph, RSS), whereas those on 2 May, 8 July and 10 July employed SALTICAM, which acts as both an acquisition camera and fast science imager. 

\subsection{NICER and Insight-HXMT}\label{sec:2.3} 

We analysed all X-ray observations of J1820 collected by NICER between March and September 2018.  NICER data were obtained from the \textit{HEASARC} data archive and processed via calibration release {\sc xti20200722}, screening out any ``hot'' or problematic detectors.

The \textit{Insight}-HXMT X-ray light-curve data collected from the \textit{Insight}-HXMT archive data centre are used here in our optical/X-ray comparative studies. The small field-of-view detectors of three payloads (LE: 1--15\,keV, 384\,cm$^2$; ME: 5--30\,keV, 952\,cm$^2$; HE: 20--250\,keV, 5100\,cm$^2$) onboard \textit{Insight}-HXMT were used to generate the light curves in good time intervals (GTIs). 

\section{Simultaneous X-ray/optical high speed photometry} \label{sec:XrayOptical}
The study of simultaneous, fast X-ray and optical variability allows us to probe the emitting regions close to the central black hole in LMXBs.  In particular, we focus on the presence of QPOs in \MAX that were detected in both X-ray and optical data.  

\subsection{QPOs} \label{sec:opticalQPOs}

We used the Lomb-Scargle (LS) Periodogram from {\tt gatspy.periodic}\footnote{{\tt Gatspy}, created by Jake VanderPlas, is a collection of tools for analyzing astronomical time series data in Python \citep{VanderPlas2015}.}, to perform the period analysis of our optical and X-ray light-curves.
 
LS power spectra of our SAAO and SALT slotmode data reveal the presence of fast variability and QPOs. We collect together all the QPOs from our SHOC and SALT data in the frequency range $\sim$0.04 Hz to 0.3 Hz in Figure \ref{fig:QPOs}, plotted in log-log space to better show the QPOs and for comparison with the X-ray behaviour. 

In Figure \ref{fig:QPOs}, the SHOC LS power spectra for days 27, 28, 89, and 93 have QPO peaks shown in red, black, green and cyan, respectively, and the blue spectrum is from SALT-RSS on day 95. The optical QPOs observed on days 27, 28, 89, 93, and 95 are at $0.072$\,Hz, $0.08$\,Hz, $0.278$\,Hz, $0.26$\,Hz, and $0.27$\,Hz, respectively, with a 1-sigma uncertainty of $0.01$\,Hz. 

\subsection{Synchronous Optical and X-ray QPOs} \label{sec:opticalXrayQPOs}

\begin{figure}
\centering
\includegraphics[width=0.46\textwidth,height=18.75pc]{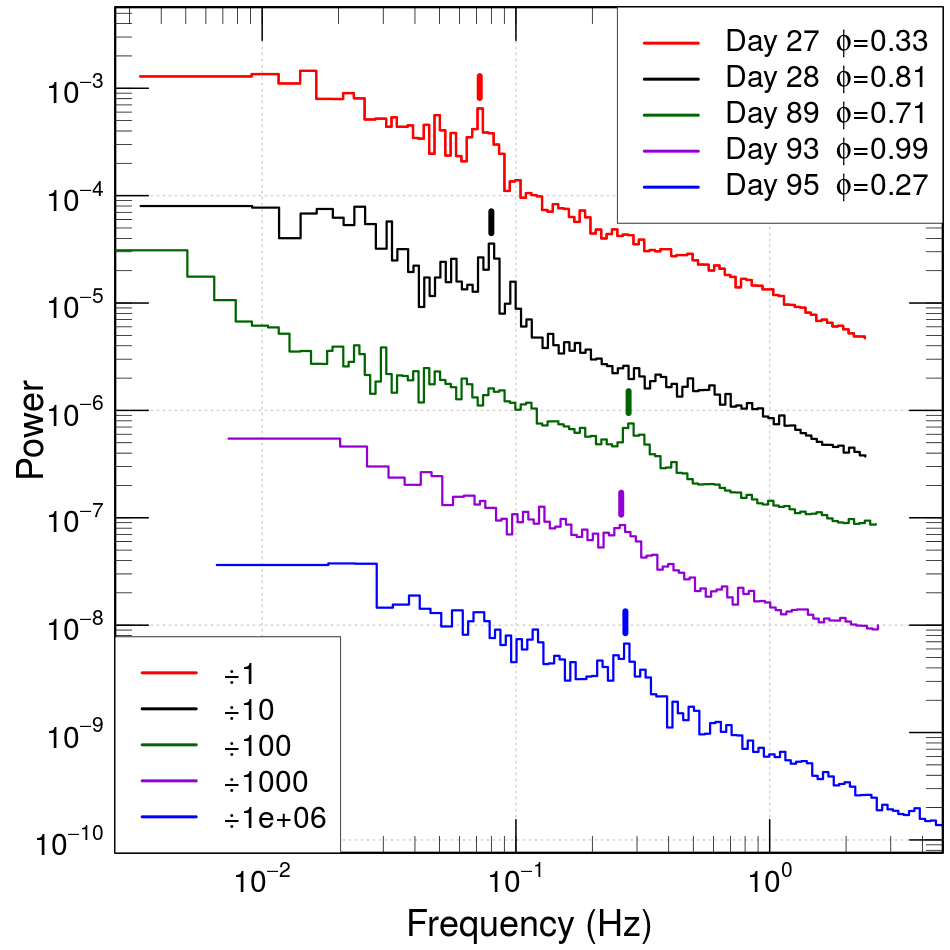}
\caption{\label{fig:QPOs}Optical QPOs from SHOC and SALT LS power spectra, identified with colour coding for each day (see box) and arbitrary offsets with power. The vertical bars mark the value of the NICER {\it X-ray} QPO fundamental frequencies \citep{Stiele2020}, apart from those on days 93 and 95, which are inferred from the upper-harmonic values.}
\end{figure}

The optical QPOs from our SHOC and SALT observations track the X-ray QPO frequencies from \cite{Stiele2020} almost perfectly, as shown by the bars marking the X-ray QPO frequencies in Figure \ref{fig:QPOs}. This immediately indicates that there must be a fundamental connection between their X-ray and optical emitting processes.

\begin{figure}
\centering
\includegraphics[width=0.475\textwidth, height=18.75pc]{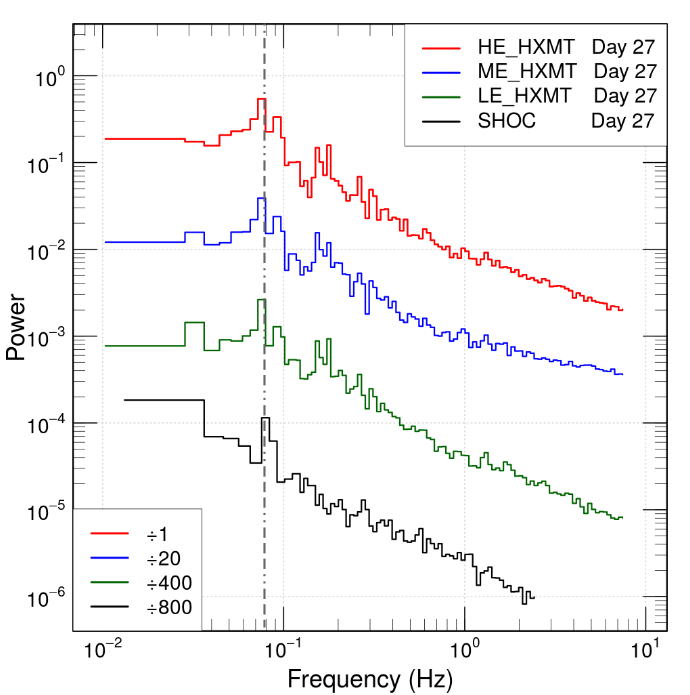}
\caption{\label{fig:HXMT} X-ray QPOs from \textit{Insight}-HXMT LS power spectra in the HE, ME and LE bands (see box for colour coding), on day 27 (7 April) simultaneous with the SHOC optical QPO. 
The vertical dotted-dashed line is passing through the QPO values. }
\end{figure}
The optical and X-ray QPOs from the simultaneous SAAO and \textit{Insight}-HXMT observations on day 27 were also analysed with LS power spectra. There too we found that the SHOC optical QPO tracks the X-ray QPOs observed from \textit{Insight}-HXMT, giving the same daily average of 0.072 Hz in both optical and X-rays. 
In Figure \ref{fig:HXMT}, we show the resulting \textit{Insight}-HXMT and SHOC power spectra when the data of day 27 are restricted to intervals of {\it exact} simultaneity (day 27 in Figure \ref{fig:QPOs} is the average power spectra for the day). This reveals that the optical and X-ray QPO features are more closely aligned (at a frequency of 0.0758 Hz) when removing any potential QPO drift between intervals.

\begin{figure}
\centering
\hspace*{-0.5cm}\includegraphics[width=0.56\textwidth,height=20pc]{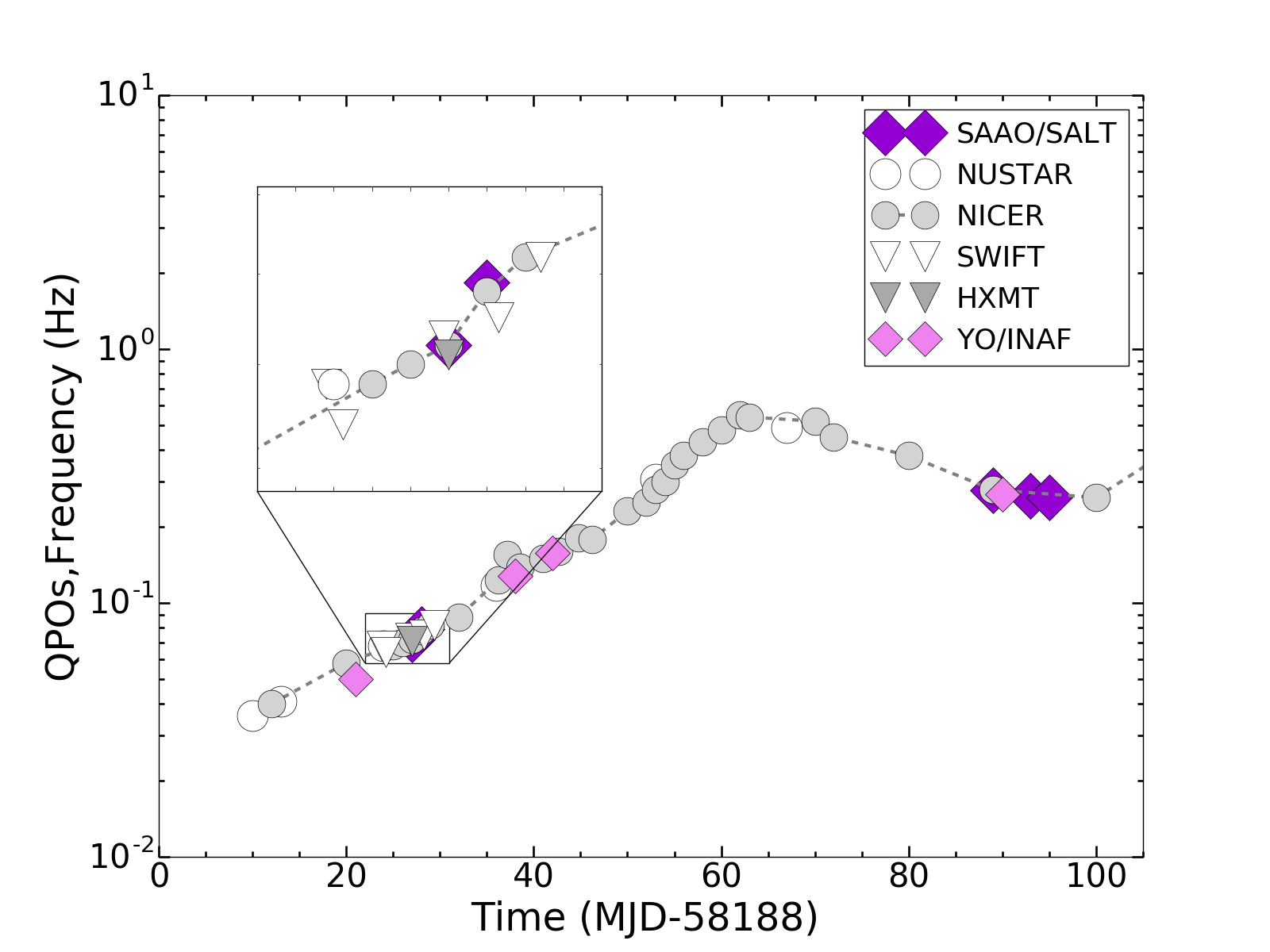}
\caption{\label{fig:QPOs_ALL} Temporal evolution of the optical and X-ray QPOs detected through the hard state of the J1820 outburst.  The grey dotted line represents a polynomial fit to the \citet{Stiele2020} NICER QPO fundamental frequencies. See box (and text) for observatory identifications.}
\end{figure}

The collection in Figure \ref{fig:QPOs_ALL} of all the X-ray and optical QPO values demonstrates how well they track each other in frequency throughout the outburst.

\subsection{Simultaneous optical/X-ray cross-correlations} \label{sec:CCF}

Optical-X-ray cross-correlations have been carried out for other bright X-ray transients (e.g. \citealt{Gandhi17} for V404~Cyg) and has already been performed on J1820 by \cite{Paice19, Paice_1820Evolution_2021}. These allow us to probe the emitting regions close to the central black hole by finding how different wavelengths lag one another within each system.

Our fast SAAO photometry was undertaken simultaneously in X-rays with NICER and \textit{Insight}-HXMT, which resulted in the Cross-Correlation Functions (CCFs)  presented in Figure \ref{fig:CCF}.
These were produced by splitting the light-curves into segments of equal length (30\,s), which were then `pre-whitened' (i.e. had the linear trend removed), and a CCF produced from each using the methodology of \citet{VenablesRipley_ModernAppliedStatistics_2002}. The mean of the CCFs was then determined, and the standard error on each bin calculated.

SAAO observations were simultaneous with NICER on days 15 and 16, and with \textit{Insight}-HXMT (LE, ME, HE bands combined) on day 27. For days 15, 16, and 27, we used 45, 22, and 47 segments (totalling 1350, 660, and 1410\,s) respectively.

\section{Discussion}
\subsection{CCF Results}
\begin{figure}
    \centering
    \includegraphics[width=0.45\textwidth]{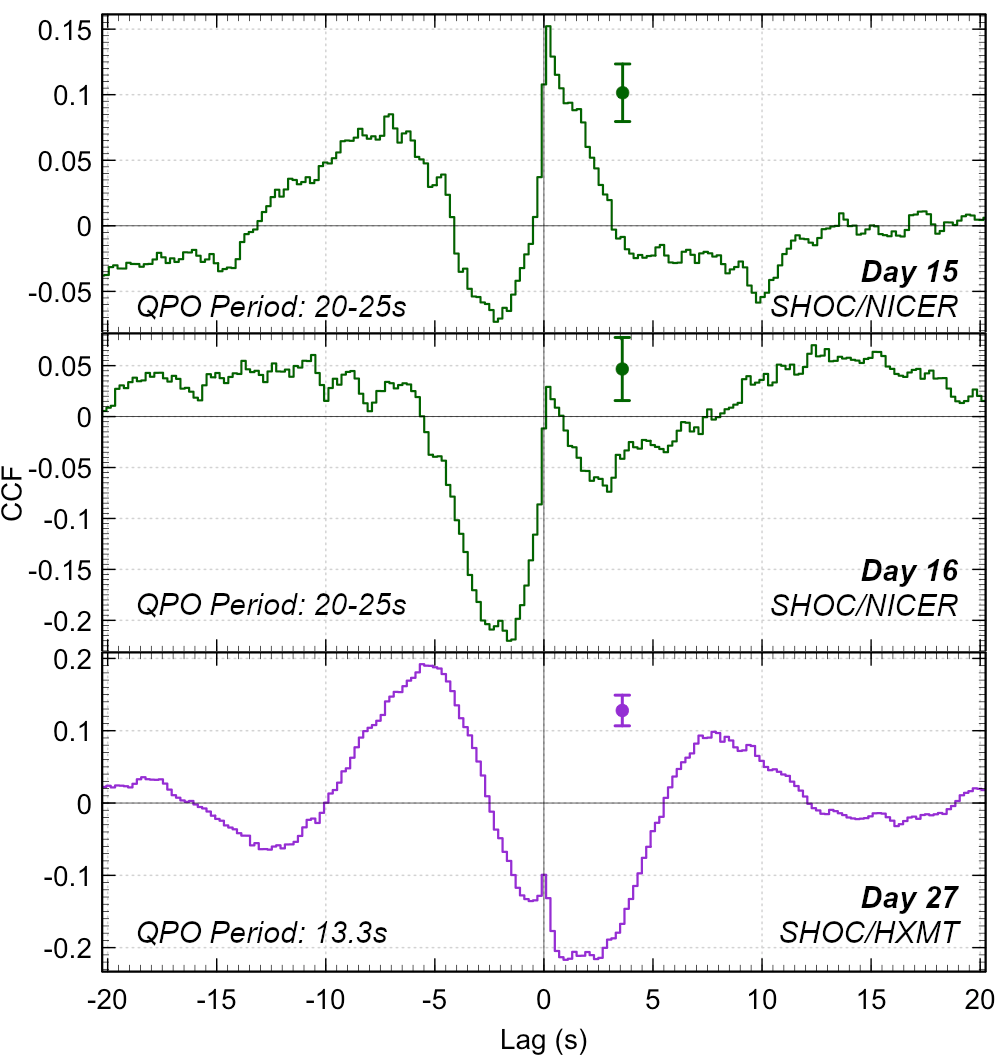}
    \caption{ Optical/X-ray CCFs for the SAAO-SHOC vs NICER data obtained on days 15 (top) and 16 (middle) (dark green), and SAAO-SHOC vs \textit{Insight}-HXMT on day 27 (bottom, dark violet). Each CCF is made from the average of 30s segments, with a representative standard error bar shown. Note the precognition dip present in days 15 and 16, which evolves into a broad anti-correlation by day 27. Also note the sub-second peak at positive lags, which changes in shape over time.}
    \label{fig:CCF}
\end{figure}

The CCFs presented in Fig. \ref{fig:CCF} show some key features also seen in other BHXBs. For example, a sub-second optical lag has also been seen in XTE\,J1118+480 \citep{kanbach_correlated_2001}, GX\,339-4 \citep{Gandhi_Correlations_2008, Gandhi2010}, and V404\,Cyg \citep{Gandhi17}. Additionally, a dramatic anti-correlation at negative lags has been seen in several sources, most significantly in Swift\,J1753.5-0127 \citep{durant_swift_2008} and, to a lesser extent, BW\,Cir \citep{pahari17} and XTE\,J1118+480.

There are multiple ways to physically interpret these features, and they have been variously ascribed to such phenomena as a jet \citep{Malzac_GX339_2018} and a hot inner accretion flow \citep{veledina_synchrotron_2011, veledina_accretion_2013}.  It is currently unclear as to which phenomena are the most significant in J1820, and what its geometry is -- see, for example, discussions on a changing coronal height \citep{Kara_MAXIJ1820Corona_2019}, a moving inner disc radius \citep{Buisson_NuSTARJ1820_1_2019, Zdziarski_GeometryJ1820_2021, DeMarco_J1820InnerFlow_2021}, the presence and strength of a jet \citep{Bright_J1820_2020, Tetarenko_J1820Jet_2021}, and some combination of jet and hot flow components \citep{Veledina_Polarisation_2019, Paice_1820Evolution_2021}. There has also been a suggestion \citep{Poutanen21} that the BH spin and orbital plane are mis-aligned by at least 40$^\circ$.

What do our CCFs show? One change between days 15 and 27 is the reduction in significance of the sub-second lag. This was also seen in \citet{Paice_1820Evolution_2021} where it was ascribed to the jet decreasing in significance; these results put constraints on the timescale over which this occurred.

Our data from day 27 show a unique feature in J1820 that has not been seen as strongly in other BHXBs; a broad anti-correlation that straddles zero lag, rather than just at negative lags. Only GX\,339-4 has shown something similar, and at much lower significance \citep{Gandhi_Correlations_2008}. Our data also show that this feature was still developing on day 16.

Another key feature is the QPO mentioned in Section \ref{sec:opticalQPOs}, which can be seen clearly in the CCF on day 27. At this point in the outburst, the QPO has a frequency of around 0.075\,Hz, which corresponds to a period of 13.3\,s, which is roughly the distance between the negative correlation peak at -5\,s and the positive correlation peak at +7\,s. This effect, of the QPO being visible in the CCF with the positive peak just over $\pi$ radians away, was also seen by \citet{Paice_1820Evolution_2021} who showed that it continued much later into the outburst (from day 37 up to at least day 88); the same work also showed the QPO to be strongly coherent. It is possible that the broad anti-correlation at positive lags could be at least partly due to the QPO, a possibility backed up by the QPO becoming more significant between our first and last CCFs (days 15 and 27; see Fig. \ref{fig:QPOs_ALL}, as well as \citealt[][Fig. 8]{Stiele2020}).

\subsection{Optical and X-ray QPOs} \label{sec:qpo_discuss}

X-ray QPOs in the mHz frequency range have been previously detected from J1820 and it is exceptional to observe type-C QPOs below 1\,Hz for large parts of a BHXB outburst \citep{Stiele2020}.

What is more remarkable in J1820 is that our optical photometry not only confirms the presence of optical QPOs in the frequency range of $\sim$0.04 Hz to 0.3 Hz from the beginning of outburst to the time of the state transition, but it also shows how well the optical and X-ray QPOs track each other throughout this time (Figure \ref{fig:QPOs_ALL}).
We therefore have the following key properties that must be explained by any physical model for J1820:

\begin{itemize}
    \item  almost perfect tracking of the optical and X-ray QPOs throughout the hard state, across five orders of magnitude in photon energy, right up to the state transition itself, 

    \item optical QPO exhibits no detectable harmonics, whereas the X-ray QPO does, suggesting the optical emitting mechanism is more sinusoidal compared to the X-ray,
    
    \item optical QPO fractional rms ($\sim$3\%) is smaller than that in X-rays ($\sim$10\%), 
    
    \item apart from the QPOs, the power spectra of both bands can be adequately fitted with two Lorentzians,

    \item \cite{Paice19} have already shown with their HiPERCAM-NICER campaign on J1820 that there is a $\sim$165ms delay between the X-ray and optical variations, together with a second, broader anti-correlated component on timescales of a few seconds. This is what we also observe. 

   \item the restricted optical/\textit{Insight}-HXMT power spectrum analysis shows that the optical is as ``sharp'' a feature as the X-ray.
\end{itemize}

Taken together, these properties indicate that the X-ray and optical QPO-emitting regions must be very close (within $\sim$0.1 lt-sec), so there is no possibility of reprocessing at the outer disc or donor star driving the optical QPO. The data are consistent with the presence of two components dominating the variability.

We note that the already known X-ray/optical delay in J1820 is actually very similar to that seen in V404 Cyg \citep{Gandhi17}, and so we should consider the scenario shown in their Figure 3. To this we can add the production of both X-ray and optical QPOs by making the jet precess at the QPO frequency, as proposed by \citet{Ma2021}. In their model, the jet precession can well explain the large soft phase lag and the energy-related behaviour of the J1820 X-ray QPO from 1--250\,keV. Furthermore, an optical QPO with a centroid frequency similar to that seen in X-rays is also expected. This has been suggested by \citet{Garcia21} in accounting for the X-ray spectral timing properties of the QPOs in MAXI~J1348-630, and is based on the MHD simulations of relativistic, precessing jets by \citet{Liska18}, as shown in their Figure 2.  

\begin{figure}
\centering
\includegraphics[width=0.46\textwidth]{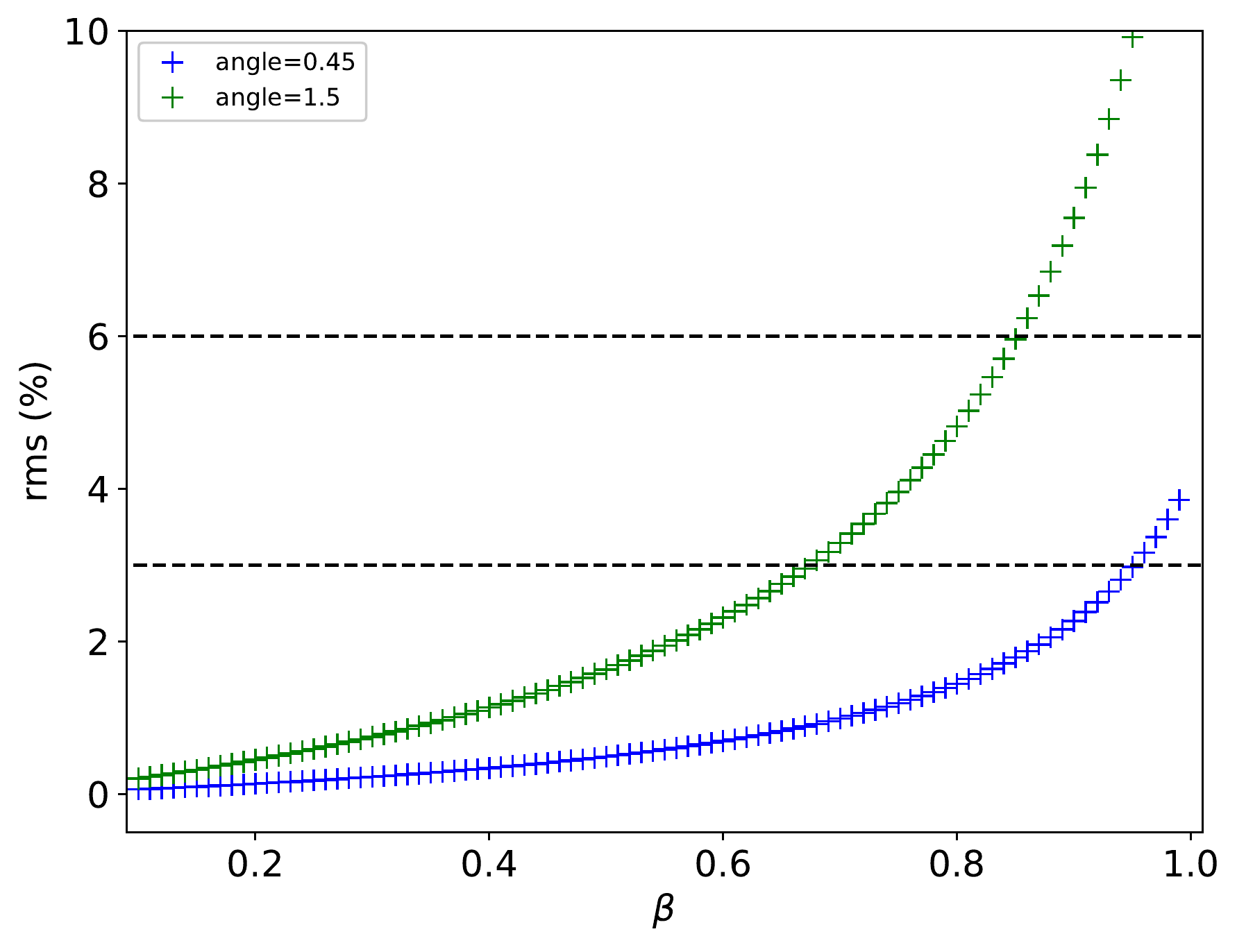}
\caption{Relation between QPO rms and  $\beta$ ($= v_{\rm jet}/c$) for opening angles of $\phi=1.5^{\circ}$ (green, \citet{Zdziarski_GeometryJ1820_2021}) and $\phi=0.45^{\circ}$ (blue, \citet{Tetarenko_J1820Jet_2021}). The lower horizontal dashed line indicates the original QPO rms of 3\%, while the upper line indicates the QPO rms of 6\% from the jet when considering the contribution of the disk emission to the optical flux \citep{Shidatsu18}. 
\label{fig:rms}  
}
\end{figure}

Additionally, if the X-ray base was more compact, then the precession would produce a more ``square'' wave, whereas the optical emitting region, higher in the jet, might be more smoothed out, explaining the different harmonic structure that we observe. Moreover, if the top of the jet is less curved, the fractional rms of the optical QPO should be smaller than the X-ray QPO, as the velocity modulation seen by the observer is relatively weak. Based on the model of \citet{Ma2021}, the optical QPO rms of 3\% would then suggest that the optical jet is faster than 0.8\,$c$ (see Figure~\ref{fig:rms}) - if we use the opening angles from the radio jet \citep{Zdziarski_GeometryJ1820_2021,Zdziarski_Jet_Para_1820_2021, Tetarenko_J1820Jet_2021}, and considering that the jet contributes about half the optical emission \citep{Shidatsu18}; the phase lag of $\pi$ indicates that the jet is curved, with the phase difference between the optical and X-ray emitting regions $\sim 180^{\circ}$.

The internal shock model of \citet{Malzac_GX339_2018} can reproduce the $\sim$\,150\,ms optical lag, as well as the anti-correlated variability through variations in Doppler-boosted jet emission. The properties of the precessing jet are driven by processes in the inner disc, and, as this moves inwards as the transition approaches, it must increase the QPO frequency. Furthermore, it will raise the X-ray base \citep{DeMarco_J1820InnerFlow_2021}, providing the raised X-ray emitting region required to explain the optical ``superhump'' (Paper 1) from the outer disc regions.

While the above is a plausible self-consistent description, it does not rule out contributions from other mechanisms. In particular, the hot flow precession model of \citet{Veledina_Poutanen_2013} predicts optical and X-ray variations that are out-of-phase with each other at inclination angles of about 60$^\circ$ (similar to J1820). But the hot flow model cannot, by itself, produce the sharp 150 ms optical lag. So a combination of a central hot flow producing the QPOs and an inner jet producing the $\sim$ 150 ms lag also appears to be viable. 

\section*{Data Availability}
The optical and X-ray data underlying this article are available from the following archives:
\begin{itemize}
    \item SALT/SAAO --
    \href{http://cloudcape.saao.ac.za/index.php/f/1009821}{http://cloudcape.saao.ac.za/index.php/f/1009821}
    \item NICER --  \href{https://heasarc.gsfc.nasa.gov/docs/archive.html}{https://heasarc.gsfc.nasa.gov/docs/archive.html}
    \item Swift -- \href{https://www.swift.ac.uk/archive/}{https://www.swift.ac.uk/archive/}
    \item \textit{Insight}-HXMT -- \href{http://hxmtweb.ihep.ac.cn/}{http://hxmtweb.ihep.ac.cn/}
\end{itemize} 
\section*{Acknowledgements}
The SALT observations were obtained under the SALT Large Science Programme on transients (2018-2-LSP-001; PI: DAHB) which is also supported by Poland under grant no. MNiSW DIR/WK/2016/07. DAHB and SBP acknowledge research support from the National Research Foundation. JPL was supported in part by a grant from the French Space Agency CNES. L.T. acknowledge research support from the National Natural Science Foundation of China. 

\bibliographystyle{mnras}
\bibliography{bibliography} 


\end{document}